\documentclass{article}
\usepackage{latexsym}
\begin{document}
\title{Unification of Gravity and Electro-Magnetism in Six Dimensions}
\author{{\bf Merab Gogberashvili}\\
Andronikashvili Institute of Physics, 6 Tamarashvili Str.,
Tbilisi 380077, Georgia \\
{\sl E-mail: gogber@hotmail.com }}
\maketitle
\begin{abstract}
Recently by us was proposed the model where Einstein's equation on the brane 
was connected with Maxwell's multi-dimensional equations in pseudo-Euclidean 
space. Based on this idea unification of 4-dimensional gravity and 
electromagnetism in (2+4)-space is found. In this picture photon is massless in 
four dimensions and obtains large mass in extra (1+1)-space normal to the brane.
\medskip 

\end{abstract}
\vskip 0.5cm

One of the main ideas of the brane models is that Plank's scale $M_{Pl}$ can be 
constructed with the fundamental scale $M$ and the brane width $\epsilon$ 
\cite{ADD}. In our previous paper \cite{v} it was shown that in the brane 
approach Einstein's equations can be effective as well. Gravitation was 
connected with the geometry of the brane embedded in pseudo-Euclidean space, 
where solutions of 4-dimensional Einstein's equation could be constructed with 
the solutions of multidimensional Maxwell's equations. This approach is the 
inverse to the standard Kaluza-Klein picture, since the vector field is 
fundamental and gravity is obtained from the brane geometry. In this direction 
geometrical unification of different interactions is easier, since Dirac 
equation also can be derived from the constrained Yang-Mills Lagrangian 
\cite{spinor}. The surface where Einstein's equations appear is not necessary 
to be 4-brane. For example, it can be 5-dimensional space with exponential warp 
factor needed to cancel cosmological constant \cite{GRS}. Embedding of this 
5-space into Euclidean (2+4)-space was found in \cite{Go}. 

The main idea of \cite{v} was based on the fact that any $n$-dimensional 
Riemannian space can be embedded into $N$-dimensional pseudo-Euclidean space 
with $n \le N \le n(n+1)/2$ \cite{EF}. This means that the multidimensional 
'tetrad' field $h^{A}_{\alpha}$ can be expressed as a derivative of embedded 
functions 
\begin{equation} \label{tetrad}
h^{A}_{\alpha} = \partial_\alpha \phi^A ~.
\end{equation}
Capital Latin letters $A,B,\dots $ numerate coordinates of embedded space, 
while Greek indices $\alpha,\beta,\dots$ numerate coordinates in four 
dimensions. 

The metric of bulk flat space-time (where the brane with arbitrary geometry is 
embedded) can be transformed to the Gaussian normal coordinates 
\begin{equation} \label{gaussian}
ds^2 = \eta_{ij}dx^idx^j + g_{\alpha\beta}(x^i,x^\nu)dx^\alpha dx^\beta ~.
\end{equation}
Here $\eta_{ij}$ is the Minkowskian metric of the normal space to the brane and 
small Latin indices $i,j,\dots$ numerates extra $(N - 4)$ coordinates. The 
induced metric on the brane is $g_{\alpha\beta}(0,x^\nu)$.

Embedding theory allows us to rewrite Hilbert's 4-dimensional action in terms 
of derivatives of the normal components $\phi^i $ of some multi-dimensional 
vector 
\begin{equation}\label{GS}
S_{g}=-M_{Pl}^2 \int R \sqrt{-g}d^4x = \eta_{ij} M_{Pl}^2\int 
\left(\Box\phi^i\Box\phi^j -
\partial_\alpha\partial_\beta\phi^i \partial^\alpha\partial^\beta\phi^j 
\right) \sqrt{-g}d^4x ~,
\end{equation}
where $\Box = \partial_\alpha \partial^\alpha$ is the 4-dimensional wave 
operator and $g$ is the determinant in Gaussian coordinates. 

In \cite{v} it was shown that to the same form (\ref{GS}) can be transformed 
Maxwell's multi-dimensional action
\begin{equation} \label{MS}
S_A = - \frac{1}{4}\int F_{AB}F^{AB} d^NX = - \frac{1}{2}\int 
[\partial_A A_B (\partial^A A^B + \partial^B A^A) - 2\partial_A 
A^A\partial_B A^B] d^NX ~,
\end{equation}
where $F_{AB} = \partial_A A_B - \partial_B A_A$ and $X^A$ are N-dimensional 
Cartesian coordinates. For this reduction we must transform $X^A$ to the 
Gaussian coordinates (\ref{gaussian}) and insert in (\ref{MS}) the exact 
solution of Maxwell's equations for the brane
\begin{eqnarray} \label{sol}
A^\alpha = \left(\frac{2}{\pi\epsilon^2}\right)^{(N-4)/4}M\sum_i^{(N-4)}
exp \left[-\frac{(x^i)^2}{\epsilon^2}\right] 
\partial^\alpha \phi^i (x^\beta ) ~, \nonumber \\
A^i = - \left(\frac{2}{\pi\epsilon^2}\right)^{(N-4)/4} \frac{2Mx^i}{\epsilon^2}
~exp \left[-\frac{(x^i)^2}{\epsilon^2}\right]\phi^i (x^\beta) ~.
\end{eqnarray}
Then we must integrate (\ref{MS}) by the extra coordinates $x^i$, remove 
boundary terms and as in \cite{ADD} put
\begin{equation} \label{M}
M_{Pl}^2 = \epsilon^{(N-4)}M^2 ~.
\end{equation}

In the present paper we want to find 4-dimensional unification of gravity and 
Electro-Magnetism based on Maxwell's equations in 6-dimensional 
pseudo-Euclidean space with the signature $(2+4)$. 

The (2+4)-space is interest object for investigations. As it was found a long 
time ago Schwarzschild's metric can be embedded in (2+4)-space \cite{ros}; also 
non-linear 15-parameter conformal group can be written as a linear 
Lorentz-type transformations there \cite{Dir}. For compact Kaluza-Klein models 
(2+4)-space was studied in \cite{PI} and in the context of brane models with 
non-factorizable geometry in \cite{GM}.

In our previous paper \cite{v} was considered only possible connection of 
Einstein's equations with multi-dimensional Maxwell's equation and was not 
received Maxwell's equations on the brane. In (\ref{sol}) there presents only 
derivatives of $\phi^i$ (which is connected with gravity) and it is clear that 
$F_{\alpha\beta}$ is zero. Now we want to insert in (\ref{sol}) also 
non-integrable part $a^\alpha (x^\beta )$ and identify it with the 
4-dimensional vector-potential. So, in (2+4)-space we wand to rewrite 
(\ref{sol}) in the form
\begin{eqnarray} \label{sol1}
A^\alpha = \sqrt{\frac{2}{\pi\epsilon^2}}
~\left[ M~ exp \left(-\frac{\tau^2}{\epsilon^2}\right)\partial^\alpha \phi^\tau 
(x^\beta ) + M~ exp \left(-\frac{\kappa^2}{\epsilon^2}\right)\partial^\alpha 
\phi^\kappa (x^\beta ) + exp \left(-\frac{\tau^2+\kappa^2}{\epsilon^2}\right) 
a^\alpha\right] ~, \nonumber \\
A^\tau = - \sqrt{\frac{2}{\pi\epsilon^2}}~\frac{2M\tau}{\epsilon^2}
~exp \left(-\frac{\tau^2}{\epsilon^2}\right)\phi^\tau (x^\beta) ~, \\
A^\kappa = - \sqrt{\frac{2}{\pi\epsilon^2}}~\frac{2M\kappa}{\epsilon^2}
~exp \left(-\frac{\kappa^2}{\epsilon^2}\right)\phi^\kappa (x^\beta) ~, 
\nonumber 
\end{eqnarray}
where $\kappa$ and $\tau$ are the normal to the brane space-like and time-like 
coordinates respectively. 

Using the anzats (\ref{sol1}) Maxwell's action (\ref{MS}) in (2+4)-space takes 
the form
\begin{eqnarray} \label{F}
S_A = - \frac{1}{4}\int F_{AB}F^{AB} d^6X = \nonumber \\
= - M^2\epsilon^2 \int R \sqrt{-g}d^4x -\frac{1}{2\pi\epsilon^2}\int exp 
\left[-\frac{2(\tau^2+\kappa^2)}{\epsilon^2}\right]
\left[ f_{\alpha\beta}f^{\alpha\beta} + \frac{4}{\epsilon^4}(\tau^2- 
\kappa^2)a_\alpha a^\alpha \right]\sqrt{-g}d^4xd\tau d\kappa = \\
= - \int \left( M_{Pl}^2 R +\frac{1}{4} f_{\alpha\beta}f^{\alpha\beta} 
\right)\sqrt{-g}d^4x ~,\nonumber
\end{eqnarray}
where $f_{\alpha\beta} = \partial_\alpha a_\beta - \partial_\beta a_\alpha$. 
Latin indices are summing with the 6-dimensional Minkowskian metric tensor 
$\eta_{AB}$ and Greek indices with the induced on the brane metric tensor 
$g_{\alpha\beta}$. So, 6-dimensional Maxwell's action on the brane is reduced 
to the 4-dimensional Einstein-Maxwell action. As in \cite{GM} the brane is 
placed at $\tau^2 = \kappa^2$ what means that it moves with the speed of light 
in transversal (1+1)-space.

In Gaussian coordinates (\ref{gaussian}) Maxwell's 6-dimensional equations 
\begin{equation}\label{Max}
\partial_AF^{AB} = 0 ~,
\end{equation}
after inserting there the anzats (\ref{sol1}) reduces to the system
\begin{eqnarray}\label{Maxwell}
\frac{1}{\sqrt{-g}}\partial_\alpha (\sqrt{-g}f^{\alpha\beta}) - 
\frac{4}{\epsilon^4}(\tau^2 - \kappa^2) a^\beta = 0 ~, \nonumber \\
\partial_\alpha a^\alpha = 0 ~.
\end{eqnarray}
One can see that the situation is similar to superconductors. On the brane 
($\tau^2 = \kappa^2$) photon is massless and obtains the large mass 
$\sim 1/\epsilon $ in the direction normal to the brane.

At the end we want to note that even if we don't want to connect $\phi^i$ with 
gravity the anzats (\ref{sol1}) can be used for the trapping of photons on the 
brane in the (2+4)-space.
\medskip

{\bf Acknowledgements:} Author would like to acknowledge the hospitality 
extended during his visits at the Abdus Salam International Centre for 
Theoretical Physics where this work was done.

\end{document}